\documentclass{article}

\usepackage[preprint, nonatbib]{neurips_2019}

\usepackage[utf8]{inputenc} 
\usepackage[T1]{fontenc}    
\usepackage{hyperref}       
\usepackage{url}            
\usepackage{booktabs}       
\usepackage{amsfonts}       
\usepackage{amsmath}
\usepackage{amssymb}
\usepackage{bbm}
\usepackage{graphicx}
\usepackage[vlined]{algorithm2e}
\usepackage{nicefrac}       
\usepackage{microtype}      

\hypersetup{
    colorlinks=true,
    linkcolor=blue,
    filecolor=magenta,      
    urlcolor=blue,
}

\title{A Survey of Online Auction Mechanism Design Using Deep Learning Approaches}

\author{%
  Zhanhao Zhang\\
  Department of Statistics\\
  Columbia University\\
  zz2760@columbia.edu
}

\begin{document}

\maketitle

\begin{abstract}
Online auction has been very widespread in the recent years. Platform administrators are working hard to refine their auction mechanisms that will generate high profits while maintaining a fair resource allocation. With the advancement of computing technology and the bottleneck in theoretical frameworks, researchers are shifting gears towards online auction designs using deep learning approaches. In this article, we summarized some common deep learning infrastructures adopted in auction mechanism designs and showed how these architectures are evolving. We also discussed how researchers are tackling with the constraints and concerns in the large and dynamic industrial settings. Finally, we pointed out several currently unresolved issues for future directions.
\end{abstract}


\section{Introduction}
Auction has been adopted as a way to negotiate the exchanges of goods and commodities for centuries. Traditionally, Generalized second price auction (GSP) and Vickrey–Clarke–Groves auction (VCG) are widely used. However, GSP is no longer a truthful mechanism if a seller has more than one item to bid. VCG is an auction mechanism based on sealed-second price auction, where winners are charged on the reductions of social welfare of other participants. Nevertheless, the VCG mechanism generates low seller revenues and does not enforce monotinicity of seller's revenues in the set of bidders and the amounts bid. It is also a non-truthful mechanism that is susceptible to multiple bids under same person or collusion of losing bidders \cite{LovelyLonelyVCG}.

Auction mechanisms with the properties of incentive compatiblility (IC) and individual rationality (IR) are highly desirable. If an auction is IC, then all bidders will truthfully reveal their private valuations of the items, so that platform administrators do not have the burden of considering bidders' strategic behaviors and are therefore able to build a reliable and predictable system. All agents are guaranteed to have non-negative utilities if the auction system is IR, and it is a very important feature that allows the system to retain its customers in the long run.

The groundbreaking work by Myerson \cite{myerson} has defined the optimal strategyproof auction for selling a single item, but limited progress has been made in characterizing strategyproof and revenue-maximizing auctions beyond this setting \cite{ProportionNet}. The dynamic nature of online auction platforms \cite{RLECommerce, RLMechanismDesign, MultipleMetricsDeepGSP} has made the problems more challenging, as the bidders, items, and platform's objectives are changing over time. In the meantime, multiple performance metrics are required to be taken into considerations in order to make the auction system attractive to bidders, sellers, and the platform \cite{NeuralAuction, RLMechanismDesign, MultipleMetricsDeepGSP}.

With the advancement of technology, researchers are shifting gears towards deep learning approaches for the design of auction mechanisms. To the best of our knowledge, the deep learning architecture is built on top of either the hybrid multiple linear perceptron infrastructure, the RegretNet infrastructure \cite{RegretNet}, the reinforcement learning infrastructure, or the DeepSet infrastructure \cite{DeepSets}. Blessed with the modern computing power, researchers are not only able to maximize the revenue, but also dealing with data sparsity and high-dimensional data, optimizing multiple performance metrics, preventing fraud, and enhancing fairness.

This article is organized as follows: we will first introduce the four common infrastructure of the deep neural networks for auction mechanism design. Then, we will discuss how researchers are tackling with the constraints and concerns other than maximizing the total revenue. Lastly, we will point out some unresolved issues and potential future directions.

\section{Hybrid Multiple Linear Perceptron (MLP) Infrastructure}
Many neural network structures are built by stacking several MLP structure into a cohesive one \cite{AutoMechanismDesign, DeepInterestNetwork, ChargingSchedulingDLMultiDrone, MobileBlockChainAuction}. The network is usually comprised of more than one components, where each components is a fully-connected feedforward neural network.

The most delicate architecture is the MenuNet \cite{AutoMechanismDesign}, which is comprised of a mechanism network and a buyer network. The mechanism takes a one-dimensional 1 as input and outputs an allocation matrix and a pricing vector. The allocation matrix contains the allocation of all items, which is obtained by a fully-connected layer followed by a sigmoid activation function. The payment vector is obtained by simply multiplying the constant 1 by a vector and it is used to represent the prices for different menu items. The buyer network doesn't require any training. It takes the outputs from the mechanism network and computes the final utility based on buyer's value profile. The training of MenuNet is very fast as the network structure is very simple. It is built upon the taxation principle \cite{taxation}, which states that simply letting the buyer do the selection can give an IC mechanism. It does not require buyer's utility function since the network only outputs buyer's strategy. It does not make any assumptions about buyer's valuation and does not require any additional constraints (such as IC and IR) to be enforced to the network. Theoretical proofs have shown that the MenuNet always return revenue optimal mechanism with IC satisfied for menus of size 2 or 3.

Zhou et al has proposed Deep Interest Network \cite{DeepInterestNetwork} for the use of click-through rate prediction. It is built on the structure of MLP but aims to conquer the bottleneck caused by the fixed-length representation vector used in traditional MLP, which has limited ability in capturing user's diverse interests from rich historical behaviors. The authors designed a local activation unit that can adaptively learn the representation of user interests from historical behaviors with respect to a certain advertisements. The data adaptive activation they adopted is Dice (Equation \ref{eq:Dice}), which is a generalization of PReLu \cite{PReLu}.
\begin{equation} \label{eq:Dice}
    f(s) = p(s) \cdot s + (1 - p(s)) \cdot \alpha s, p(s) = \frac{1}{1 + e^{-\frac{s - E[s]}{\sqrt{Var[s] + \epsilon}}}}
\end{equation}

In the training phase, $E[s]$ and $Var[s]$ are the mean and variance of input in each-minibatch, while in the testing phase, $E[s]$ and $Var[s]$ are moving averages of $E[s]$ and $Var[s]$ over data. $\epsilon$ is a small constant and is set to be $10^{-8}$ by authors.

They adopt different representation vector for different ads. The embedding layer uses single embedding vector (one-hot) and multiple embedding vectors (multi-hot) in combination. Pooling and concat layers are added to transform the list of embedding vectors into the same lengths, so that the network can allow different users to have different number of behaviors.

Shin et al transformed the charging scheduling problem into an auction problem using deep learning framework \cite{ChargingSchedulingDLMultiDrone}. It is designed based on the concept of the Myerson auction \cite{myerson}, which is one of the most efficient revenue-optimal single-item auctions. One of the most challenging issues about charging scheduling is the lack of prior knowledge on the distribution of the number of bidders. Employing the auction approach is useful when there is no accurate information of buyer's true valuation, and buyers are not aware of the private true values of other buyers. Buyer's values are represented by the urgency of drone machines, while seller's revenue is generated from the payment from resource allocation. As Myerson's auction system requires full knowledge of the distribution of bids in order to compute the expected payment, Shin et al used deep neural network to parametrize the virtual valuation function, the allocaton rule, and the payment rule.

The network begins with a monotonic network that transforms the bids $b_i$ of the drone into $\bar{b}_i$ using the virtual valuation function $\phi^{mononet}$ parametrized by the network. In the $\phi^{mononet}$, all outcomes of $\phi^{shared}$ (Equation \ref{eq:Mono2}) are computed using the same weights, while the $\phi_i$ (Equation \ref{eq:Mono1}) calculates the outcome using different weights for each bid.

\begin{equation} \label{eq:Mono1}
    b_i' = \phi_i(b_i) = \min_{1 \leq g \leq \mathcal{G}} \{\max_{1 \leq n \leq \mathcal{N}} (w_{g, n}^i b_i + \beta_{g, n}^i)\}
\end{equation}
\begin{equation} \label{eq:Mono2}
    \bar{b}_i = \phi_i^{shared}(b_i') = \min_{1 \leq g \leq \mathcal{G}} \{\max_{1 \leq n \leq \mathcal{N}} (w_{g, n}^{shared} b_i' + \beta_{g, n}^{shared})\}
\end{equation}

The payment rule network takes in the output $\bar{b}_i$ from the monotonic network and returns $\bar{p}_i$ according to Equation \ref{eq:Payment1}. Then, the payment value is computed using the inverse function of $\phi^{mononet}$ (Equation \ref{eq:Payment2}, \ref{eq:Payment3}). Finally, the allocation rule network assigns the highest winning probability to the highest bidder with positive transformed bid.

\begin{equation} \label{eq:Payment1}
    \bar{p}_i = ReLU\{\max_{j \neq i}(\bar{b}_i)\}
\end{equation}
\begin{equation} \label{eq:Payment2}
    p_i' = \phi_{shared}^{-1}(\bar{p}_i) = \max_{1 \leq g \leq \mathcal{G}}\{\min_{1 \leq n \leq \mathcal{N}}(w_{g, n}^{shared})^{-1}(\bar{p}_i - \beta_{g, n}^{shared})\}
\end{equation}
\begin{equation} \label{eq:Payment3}
    p_i = \phi_i^{-1}(p_i') = \max_{1 \leq g \leq \mathcal{G}}\{\min_{1 \leq n \leq \mathcal{N}}(w_{g, n}^i)^{-1}(p_i' - \beta_{g, n}^i)\}
\end{equation}

Luong et al constructed a neural network \cite{MobileBlockChainAuction} for edge computing resource management based on analytical solution \cite{myerson}, which guarantees the revenue maximization while ensuring the IC and IR. The network structure also has three key components as in \cite{ChargingSchedulingDLMultiDrone}: neural network parametrized monotone transformation functions that map bids into transformed versions, an allocation rule that maps the transformed bids to a vector of assignment probabilities, and a conditional payment rule that is based on the maximum non-negative transformed bids. The allocation and payment rules are derived from SPA-0, second price auction with 0 reserve price, where the reserve price is the mininum price a seller is willing to accept from the buyer.

RochetNet \cite{RegretNet} proposed by Dutting et al. is also an application of MLP. The RochetNet is a single-layered neural network that takes in the bids and outputs the maximum non-negative transformed values. It is used to model a non-negative, monotone, convex, and Lipschitz utility function, using $J$ linear functions with non-negative coefficients. The RochetNet easily extends to a single bidder with a unit-demand valuation \footnote{Unit-demand valuation: the value of a subset is the maximum individual valuation within that subset.}. Each linear function in the RochetNet corresponds to an option on the menue, with the allocation probabilities and payments encoded through its slope and intercept.

The MoulinNet \cite{DLMultiFacility} proposed by Golowich et al. also adopts the structure of MLP, which is used to determine the optimal facility locations preferred by agents. MoulinNet is a monotone feed-forward neural network that learns the generalized median rules \cite{Moulin}. For single-facility mechanisms, the mechanism in Equation \ref{eq:Moulin} is strategy-proof, which selects the median of agents' most preferred locations (the agents' peaks). The inputs of the network are binary-encoded vectors $\nu(S)$ that represent whether the bidded items in $S$ are selected. $w$ and $b$ are parameters in MoulinNet. The $u_i$ is the utility function for agent $i$ and $\tau$ represents the peaks of the facility. The output of the network is the optimal selection rules based on utilities.
\begin{equation} \label{eq:Moulin}
    f^{w, b}(u) = \min_{S \subseteq N} \{\max_{i \in S}\{\tau(u_i), h^{w, b}(\nu(S))\}\}
\end{equation}

\section{RegretNet Infrastructure}
The RegretNet \cite{RegretNet} proposed by Dutting et al is comprised of an allocation network and a payment network. Both are built upon the MLP infrastructure, but the RegretNet has been adopted and extended in the auction designs in various settings \cite{DLWithBudgets, DLMultiFacility, PreferenceNet, ProportionNet}.

Two basic assumptions are required by the RegretNet architecture: additive valuation \footnote{Additive valuation: an agent's valuation for a subset of items is the sum of the individual items' valuations.} and unit-demand valuation. Both of the allocation network and the payment network takes in the bids as inputs, feeds them into MLP-structured networks with separate parameters, and returns the total payments based on the outputs from two networks. Therefore, the two networks are trained together. The network uses a sigmoidal unit to normalize the payment vector into [0, 1], so that the IR constraint will be enforced, where bidders are never charged for more than their expected value for the allocation. 

The objective function is aiming to minimize the empirical loss (negated revenue) subject to the IC and IR constraints. The IC constraint can be enforced by the notion of ex post regret for bidders, which is the maximum increase in their utility considering all possible non-truthful bids. The ex post regret is estimated by the empirical regret, which is denoted as $\hat{rgt}_i(w)$. Therefore, the objective function becomes (Equation \ref{eq:RegretObj1}):

\begin{equation} \label{eq:RegretObj1}
    \begin{split}
        \min_{w \in \mathbb{R}^d} -\frac{1}{L} \sum_{l = 1}^{L}\sum_{i = 1}^n p_i^w(v^{(l)})\\
        \text{s.t. } \hat{rgt}_i(w) = 0, \forall i \in N.
    \end{split}
\end{equation}

The optimization is achieved using Lagrange multipliers, augmented with a quadratic penalty term for violating the constraints (Equation \ref{eq:RegretObj2}):

\begin{equation} \label{eq:RegretObj2}
    \mathcal{C}_{\rho}(w; \lambda) = -\frac{1}{L} \sum_{l = 1}^{L}\sum_{i = 1}^n p_i^w(v^{(l)}) + \sum_{i \in N} \lambda_i \hat{rgt}_i(w) + \frac{\rho}{2}\sum_{i \in N} (\hat{rgt}_i(w))^2
\end{equation}

Feng et al constructed a neural network \cite{DLWithBudgets} built upon the structure of RegretNet, which consists of an allocation network and a payment network. It extends the RegretNet infrastructure by incorporating the budget constraints as well as handling Bayesian Incentive Compatible (BIC) \footnote{Bayesian Incentive Compatible: truth-telling is the optimal strategy for a bidder in expectation with respect to the types of others, given that the other bidders report truthfully.} and conditional IC constraints. Dutting et al enforces IC by requiring the empirical ex post regret to be zero, while Feng et al are able to handle more general forms of IC by constructing an appropriate notion of regret. 

Assume we have an auction with rules $(a, p)$. To handle BIC, Feng et al constrain the empirical interim regret $rgt_i(a, p)$ (Equation \ref{eq:FengRegret}) to zero. To handle conditional IC/BIC, they constrain the empirical conditional regret to zero. They also incorporate the individually rationality (IR) $irp_i(a, p)$ (Equation \ref{eq:FengIR}) and budget constraint (BC) $bcp_i(a, p)$ (Equation \ref{eq:FengBC}) as penalties.

\begin{equation} \label{eq:FengRegret}
    rgt_i(a, p) = E_{t_i \sim F_i} [\max_{t_i' \in \mathcal{T}_i} 1_{\mathcal{P}_i(t_i') \leq b_i} (\mathcal{U}_i(t_i, t_i') - \mathcal{U}_i(t_i, t_i))]
\end{equation}

\begin{equation} \label{eq:FengIR}
    irp_i(a, p) = E_{t_i \sim F_i} [\max\{0, -\mathcal{U}_i(t_i, t_i)\}]
\end{equation}

\begin{equation} \label{eq:FengBC}
    bcp_i(a, p) = E_{t_i \sim F_i} [\max\{0, \mathcal{P}_i(t_i) - b_i\}]
\end{equation}

The loss function is the negated expected revenue $\mathcal{L}(a, p) = -E_{t \sim F}[\sum_{i = 1}^n p_i(t)]$. Let $w \in \mathbb{R}^d$ denote the parameters of the allocation network, the induced allocation rule denoted by $a^w$, and $w' \in \mathbb{R}^{d'}$ denote the parameters of the payment network, the induced payment rule is denoted by $p^{w'}$. The objective function is finally in Equation \ref{eq:FengObj}. The objective function is trained using Augmented Lagrangian Solver as in Dutting et al, where the quadratic penalty terms are added for each constraint.

\begin{equation} \label{eq:FengObj}
    \begin{split}
        \min_{w \in \mathbb{R}^d, w' \in \mathbb{R}^{d'}} \mathcal{L}(a^w, p^{w'})\\
        \text{s.t. } rgt_i(a^w, p^{w'}) = 0, \forall i \in [n]\\
        irp_i(a^w, p^{w'}) = 0, \forall i \in [n]\\
        bcp_i(a^w, p^{w'}) = 0, \forall i \in [n]\\
    \end{split}
\end{equation}

Golowich et al proposed RegretNet-nm \cite{DLMultiFacility} that is able to give general mechanisms that are not limited by existing characterization results for multi-facility location problems. The notion of regret is extended to facility location mechanisms as the maximum expected utility gain agents can achieve by misreporting their preferences. 

The network structure is the same as RegretNet, except for the inputs which are agents' peaks. The misreported peaks are sampled uniformly within $[0, 1]$, with a granularity of $\frac{1}{M}$. The ex post regret is integrated into the objective function using Augmented Lagrangian Solver, which uses a quadratic penalty term. The RegretNet-nm opens the door for mechanisms designs for settings without money, such as matching and allocation problems, using neural network approaches.

PreferenceNet \cite{PreferenceNet} is another extension of RegretNet. It encodes human preferences in auction designs. The network structure is comprised of RegretNet and a 3-layer MLP. These two components are trained in an EM-manner: MLP is first trained using a uniformly drawn sample of allocations as inputs, and it is optimized using binary cross entropy loss based on ground truth labels. Then, the RegretNet is trained using Augmented Lagrange Solver. 

The loss function for the entire PreferenceNet is defined in Equation \ref{eq:PreferenceObj}, where $\mathcal{L}_s$ is the output of the trained MLP. Lastly, the allocations and payments are sampled every $c$ epochs from the partially trained RegretNet and use them to augment the MLP training set to adapt to the distributional shifts in allocations during training.

\begin{equation} \label{eq:PreferenceObj}
    \begin{split}
        \mathcal{C}_{\rho}(w; \lambda) = -\frac{1}{L} \sum_{l = 1}^L\sum_{i \in N} p_i^w(v^{(l)}) + \mathcal{L}_{rgt} - \mathcal{L}_{s}\\
        \text{Where } \mathcal{L}_{rgt} = \sum_{i \in N} \lambda_{(r, i)} \text{rgt}_i(w) + \frac{\rho_r}{2}(\sum_{i \in N} \text{rgt}_i(w))^2, \mathcal{L}_s = \sum_{j \in M} s_j
    \end{split}
\end{equation}

Peri et al. proposed Preference Classification Accuracy (PCA) metric to evaluate how well a learned auction model satisfies an arbitrary constraint. PCA is calculated by the fraction of test bids that satisfy the ground truth constraint. Then, authors use pairwise comparisons between allocations to elicit preferences. Each input set of allocations is compared against n other allocations on their preference scores and label it as either a positive or negative exemplar, based on if its preference score is higher than the majority of others.

ProportionNet \cite{ProportionNet} proposed by Kuo et al. is also based on the infrastructure of RegretNet. Like most other neural networks that deal with auction mechanism designs, it does not work under the setting of combinatorial valuations. Under the assumption of additive valuations and unit-demand valuations, the input space of valuations reduces from $2^M$ to $M$, where $M$ is the number of items. 

\begin{equation} \label{eq:KuoObj}
    \mathcal{C}_{\rho}(w; \lambda) = -\frac{1}{L}\sum_{l = 1}^L\sum_{i \in N} p_i^w(v^{(l)}) + \mathcal{L}_{rgt} + \mathcal{L}_{unf}
\end{equation}

Kuo et al. follows the core idea of RegretNet: in the Bayesian auction setting, one knows the valuation distribution from which samples can presumably be drawn. In the meantime, as both of the allocation and payment rules are functions, we can parametrize them using neural networks. Strategyproofness can be enforced adding constraints that are solvable using Augmented Lagrange Optimizer. 

It adopted same neural network architecture as RegretNet, but adding a constraint of unfairness in the loss function \ref{eq:KuoObj}, so that discriminatory ad allocations among different demography can be mitigated. The regret term $\mathcal{L}_{rgt}$ is consistent with the definition in RegretNet. The term $\mathcal{L}_{unf}$ is for quantifying the unfairness and discrimination in the auction system, which will be described in more details in Section \ref{sec:Fairness}.

\section{Reinforcement Learning (RL) Infrastructure}
As the online auction is more often a dynamic system whose users and platform objectives are evolving over time, researchers are more inclined to use dynamically trained models to adapt to the current status quo, leveraging reinforcement learning infrastructures \cite{RLECommerce, RLMechanismDesign, MultipleMetricsDeepGSP}.

As in the RegretNet infrastructure, Cai et al. also adopted no-regret learning, in which case agents only need to reason about their own strategies and their interaction with the environment, while they don't have to know the values of competitors or compute payoff-maximizing strategies over a long sequence of rounds. Reasoning about the strategies of other parties usually require strong cognitive assumption and highly burdensome computing power, which most agents don't have access to.

Based on well-known bandit algorithms, Cai et al. identified four possible strategies for sellers.
\begin{enumerate}
    \item $\epsilon-$Greedy \cite{EpsilonGreedy}: With probability $\epsilon$, each seller selects a strategy uniformly at random. With probability $1 - \epsilon$, the strategy with the best observed empirical mean payoff is selected.
    \item $\epsilon-$First: For a horizon of $\mathcal{T}$ rounds, the seller picks a strategy uniformly at random for the first $\epsilon \cdot \mathcal{T}$ rounds, and then picks the strategy that maximizes the empirical mean of the observed rewards for all the subsequent rounds.
    \item Exponential-weight Algorithm for Exploration and Exploitation (Exp3) \cite{Exp3-1, Exp3-2}: In short, Exp3 selects a price according to a weighted distribution and then adjust the weights based on payoffs. To be more precise, suppose there are $K + 1$ possible prices, the probability distribution $\pi_i(t)$ of those prices at round $t$ is defined in Equation \ref{eq:Exp3Dist}, where $w_i(t)$ are the current weight of $i^{th}$ price at round $t$ and $\gamma$ is a real number between $[0, 1]$.
    \begin{equation} \label{eq:Exp3Dist}
        \pi_i(t) = (1 - \gamma)\frac{w_i(t)}{\sum_{j = 1}^{K + 1}w_j(t)} + \gamma \frac{1}{K + 1}
    \end{equation}
    Select a price $p_j(t)$ according to the distribution above and compute its payoff $u_j(t)$, then the weight for $j^{th}$ price is updated according to Equation \ref{eq:Exp3Weights}, while the weights for all the other prices remain unchanged.
    \begin{equation} \label{eq:Exp3Weights}
        w_j(t + 1) = w_j(t)e^{\gamma\frac{u_j(t)}{(K + 1)\pi_j(t)}}
    \end{equation}
    \item Upper confidence Bound Algorithm (UCB1) \cite{UCB1-1, UCB1-2}: In the first $K + 1$ rounds, select a price not used before from $[0, \frac{1}{K}, \dots, 1]$ and then select the price with the max weighted value in the subsequent rounds. 
    
    Initialize the weights for all prices to be 0. For any round $t \in \{0, \dots. K\}$, the seller chooses a price $p_j$ such that $p_j \in [0, \frac{1}{K}, \dots, 1]$, computes the utility $u_j(t)$, and updates the weights for $j^{th}$ price to be $x_j(t) = \frac{x_j(t - 1) + u_j(t)}{t}$, and keeping the weights for all the other prices unchanged.
    
    For any round $t \geq K + 1$, the seller chooses the price $p_j$ according to Equation.
    \begin{equation} \label{eq:UCB1Price}
        p_j(t) = argmax_{j \in \{0, \frac{1}{K}, \dots, 1\}} x_j(t) + \frac{\log_2 t}{\sum_{\tau = 1}^t 1_{\{p_j \text{ was chosen in round }\tau\}}}
    \end{equation}
\end{enumerate}
$\epsilon-$First and $\epsilon-Greedy$ have a clear distinction between exploration and exploitation and belong to the class of semi-uniform strategies. Exp3 makes no distributional assumptions about the rewards and is widely used for the full information setting and works in the adversarial bandit feedback model \cite{Exp3-2}. UCB1 maintains a certain level of optimism towards less frequently played actions and uses the empirical mean of observed actions to choose the action in the next round. UCB1 is best suited in scenarios where rewards follow some unknown distributions \cite{RLECommerce}.

These sellers' models can model sellers with different degrees of sophistication or pricing philosophies, and it is consistent with the recent literature on algorithmic mechanism deisgn, in terms of modeling agetn rationality in complex dynamic environments.

Previous researchers have already come up with a few variants of reinforcement learning models. The disadvantage of Deep Q-Network (DQN) \cite{DQNShortcomings} is that it cannot handle continuous actions or high-dimensional action spaces, as stochastic actor-critic algorithms are hard to converge. The Deterministic Policy Gradient (DPG) algorithm \cite{DPG} is developed to train a deterministic policy with parameter vector. The DPG consists of the critic and actor. The critic approximates the action-value function, while the actor adjusts the parameters of the deterministic policy. Deep Deterministic Policy Gradient (DDPG) \cite{DDPG} is then proposed, as DPG is severely impacted by the high degree of temporal correlation that introduces high variance. DDPG stores the experiences of the agetn at each time step in a replay buffer and uniformly samples mini-batch from it at random for learning, which can eliminate the temporal correlation. DDPG also employs target networks for the regularization of the learning algorithm, which updates the parameters at a slower rate.

However, the size of the action space blows up very sharply with the number of sellers increases, so an direct application of DDPG will fail to converge. In addition, the DDPG is not able to handle variability on the set of sellers, since the algorithm uses a two-layer fully connected network and the positions of each seller plays an important role \cite{RLECommerce}.

Cai et al. proposed IA(GRU) \cite{RLECommerce} algorithm that aims to mitigate the problems from DDPG. It adopted the framework of DDPG by maintaining a sub-actor network and a sub-critic network. In each step of training, if utilizes a background network to perform a permutation transformations by ordering the sellers according to certain metrics, which maintains permutation invariance. In the meantime, it applies a recurrent neural network (RNN) on the history of sellers. The outputs from the permutation transformation and the outputs from the RNN on histories are then integrated together as inputs to the sub-actor and sub-critic networks.

In reality, participants of online auctions are constrained from both informational and computational aspects and therefore they are not fully rational. In addition, the historical data can be limited to the ones generated by mechanisms that are defined by only few sets of parameters, and therefore we do not have enough exploration for the past data. Both participants and auction system designers are impacted by multiple and complicated factors, and therefore their decisions are changing over time. To overcome those difficulties, Tang et al. models each player as an independent local Markov decision process \cite{RLMechanismDesign}, where a local state encodes the part of historical actions and outcomes that the player can observe so far.

Tang et al. uses DDPG infrastructure to handle continuous action space, but it decomposes the original neural netowrk into a set of sub-networks, one for each seller, to handle the huge number of states. It depends on the assumption that sellers are independent and the Q-values are additive among multiple sellers. It uses LSTM to adaptively learn from its past bidding data and feedback to predict future bid distribution, while it does not explicitly model each advertiser's bidding strategy. In order to optimize the designer's markov decision process, it discretizes the action space and then use the Monte-Carlo tree search (MCTS) \cite{MCTS} to speed up the forward-looking search. Experiments and case studies show that the dynamic pricing scheme proposed by Tang et al. outperforms all static schemes with large margins.

Another challenge in the dynamic online auction systems is the variety of performance metrics users and designers consider when making their decisions, while most state-of-the-art auction mechanisms only optimizes a single performance metrics, such as revenue or social welfare. Zhang et al. identified a list of performance metrics \cite{MultipleMetricsDeepGSP} that can be considered by users, advertisers, and the ad platform.
\begin{enumerate} \label{Metrics}
    \item Revenue Per Mille (RPM): $RPM = \frac{\sum click \times PPC}{\sum impression} \times 1000$, where PPC is the payment for winning ads.
    \item Click-Through Rate (CTR): $CTR = \frac{\sum click}{\sum impression}$.
    \item Add-to-Cart Rate (ACR): $ACR = \frac{\sum add-to-cart}{\sum impression}$.
    \item Conversion Rate (CVR): $CVR = \frac{\sum order}{\sum impression}$.
    \item GMV Per Mille (GPM): $GPM = \frac{\sum merchandise volume}{\sum impression} \times 1000$.
\end{enumerate}

Zhang et al. proposed the Deep GSP \cite{MultipleMetricsDeepGSP} that can optimize multiple performance metrics as in Equation \ref{eq:DeepGSPObj}, where $b$ is the bid vector from users, $\mathcal{M}$ is the auction mechanism, $f_j$ is the $j^{th}$ performance metrics function, and $w_j$ is the weights associated with $j^{th}$ performance metrics and can be adjusted by the auction platform administrators from time to time.
\begin{equation} \label{eq:DeepGSPObj}
    \begin{split}
        \mathcal{M} = argmax_{\mathcal{M}} E_{b \sim \mathcal{D}} [\sum_{j = 1}^L w_j \times f_j(b; \mathcal{M})]\\
        \text{s.t. Game Equilibrium constraints,}\\
        \text{Smooth Transition constraints}
    \end{split}
\end{equation}

The Deep GSP auction is built upon the classical generalized second-price auction. It takes in the features of items (e.g. category, historical click-through rate), user profile (e.g. gender, age, income), and user preference (e.g. budget, marketing demands) as inputs to a deep neural network, which integrates those features with the bids into an input vector $x_i$ and map them to a rank score $r_i = R_{\theta}(b_i; x_i)$, where $R_{\theta}(b_i; x_i)$ is the mapping function. Bidders are then sorted based non-increasingly based on their rank scores, and the top-K bidders would win the auction. The payments of the winning bidders are based on the bids from the next highest bidders.

The mapping function $R_{\theta}(b_i; x_i)$ needs to be monotone with respect to the bids $b_i$, in order to satisfy the game equilibrium constraint. Some pieces of previous research enforced monotonicity by designing specific neural network architectures, but it increases the computational complexity for the training procedure. Therefore, Zhang et al. directly incorporate the monotonicity constraint by introducing a point-wise monotonicity penalty term (Equation \ref{eq:DeepGSPMono}) into the loss function, where $\pi_{\theta}(b_i; x_i)$ is a non-linear function with bid and is parametrized using a deep neural network.

\begin{equation} \label{eq:DeepGSPMono}
    \begin{split}
        \mathcal{L}_{mono} =& \sum_{i = 1}^N \max(0, -\triangledown_b R_{\theta}(b_i; x_i))\\
        =& \sum_{i = 1}^N \max(0, -(\pi_{\theta}(b_i; x_i) + b_i\triangledown_b\pi_{\theta}(b_i; x_i) ))
    \end{split}
\end{equation}

The smooth transition constraint is imposed in Equation \ref{eq:DeepGSPSmooth}, which ensures that the advertiser's utility would not fluctuate too much when the auction mechanism is switched towards another objective.

\begin{equation} \label{eq:DeepGSPSmooth}
    u_i(\mathcal{M}) \geq (1 - \epsilon) \times \bar{u}_i(\mathcal{M}_0)
\end{equation}

\section{DeepSet Infrastructure}
Most deep neural network has an implicit constraint on the position of each input, while in reality the items in auction do not have an inherent ordering. Cai et al. uses permutation transformation to mitigate the position effect by ordering the sellers based on some metrics. Liu et al. took a step further by removing the ordering effect completely. They introduced Deep Neural Auction (DNA) \cite{NeuralAuction} that is built on the DeepSets \cite{DeepSets} architecture.

The set encoder for DNA is composed of two groups of layers $\phi_1$ and $\phi_2$. Each instance $x_i$ is first mapped to a high-dimensional latent space using the shared fully connected layers $\phi_1$, followed by the Exponential Linear Unit (ELU) \cite{ELU} activation function $\sigma$. Then, it is processed with symmetric aggregation pooling (e.g. avgpool) to build the final set embedding $h_i'$ for each ad $i$ with another fully connected layer $\phi_2$. The entire procedure is described in Equation \ref{eq:DNAPermutation}, where $h_{-i}$ represent the hidden states from all bidders except $i$.

\begin{equation} \label{eq:DNAPermutation}
    \begin{split}
        h_i = \sigma(\phi_1(x_i))\\
        h_i' = \sigma(\phi_2(\text{avgpool}(h_{-i})))
    \end{split}
\end{equation}

DNA uses context-aware rank score uses a strictly monotone neural network with respect to bid, and supports efficient inverse transform given the next highest rank score. The rank score can be obtained using Equation \ref{eq:DNARank} and the price can be obtained using Equation \ref{eq:DNAPrice}, where $w_{qz}$, $w_{qz}'$, and $\alpha_{qz}$ are weights of the neural network.
\begin{equation} \label{eq:DNARank}
    r_i = \min_{q \in [Q]}\max_{z \in [Z]}(e^{w_{qz}} \times b_i + w_{qz}' \times x_i' + \alpha_{qz})
\end{equation}

\begin{equation} \label{eq:DNAPrice}
    p_i = \max_{z \in [Z]}\min_{q \in [Q]} e^{-W_{qz}} (r_{i + 1} - \alpha_{qz} - w_{qz}' \times x_i')
\end{equation}
This partially monotone MIN-MAX neural network represented by Equation \ref{eq:DNARank} has been proved to be able to approximate any function \cite{MINMAX}.

\begin{equation} \label{eq:DNANeuralSort}
    \hat{M}_r[k, :] = \text{softmax}(\frac{(N + 1 - 2k)r - A_r\mathbbm{1}}{\tau})
\end{equation}

The DNA also model the whole process of allocation and payment inside the neural network framework, as treating allocation and payment as an agnostic environment can limit the deep learning results. One of the challenges is that both the allocation and payment are built on a basic sorting operation, which is not differentiable. Liu et al. overcome this issue by proposing a differentiable sorting engine that caters to the top-K selection in the multi-slot auctions, leveraging Neural-Sort \cite{NeuralSort}. 

In Equation \ref{eq:DNANeuralSort}, the intuitive interpretation of $\hat{M}_r[k,:]$ is the choice probabilities on all elements for getting the $k^{th}$ highest item. Where $A_r[i, j] = |r_i - r_j|$, and $\mathbbm{1}$ denotes the column vector of all ones. The top-K payments can therefore by recovered by a simple matrix multiplication in Equation \ref{eq:DNANeuralPay}.

\begin{equation} \label{eq:DNANeuralPay}
    f_{pay} = \hat{M}_r[1:K,:] \cdot [p_1, p_2, \dots, p_N]^T
\end{equation}

\section{Constraints and Concerns}
Online auction system is complex gaming system among bidders, sellers, and auction system administrators. Merely considering the revenue or social welfare for one party is more often sub-optimal. To the best of our knowledge, we have identified four large categories of constraints and concerns when administrators are designing their auction systems: IC \& IR, data sparcity \& high dimensionality, multiple performance metrics and objectives, and fairness \& fraud prevention. We will illustrate below why these constraints matter and then summarize how current researchers are tackling with them.

\subsection{Incentive Compatibility (IC) \& Individual Rationality (IR)}
Auction platform designers are more often interested in maximizing long-term objectives \cite{RLMechanismDesign}. Therefore, building a reliable system that can adapt to the dynamic and complicated environment is crucial for the success. 

IC property ensures that all agents will achieve the best outcome by reporting their values truthfully. Auction participants may come from many different background and therefore informational, cognitive, and computational constraints will limit their rationality in different extent. The stability and reliability of the system will be much harder to maintain if IC cannot be satisfied, as designers and agents have to take all potential strategic behaviors of all other agents into account. Some researchers accomplished IC by adopting theoretical frameworks, such as GSP by DNA \cite{NeuralAuction} and taxaction principal \cite{taxation} by MenuNet \cite{AutoMechanismDesign}. Other researchers \cite{RegretNet, DLWithBudgets, DLMultiFacility, ProportionNet, PreferenceNet} achieved IC by enforcing ex post regret \cite{RegretNet} to zero.

IR property is also important as it ensures that all agents are receiving non-negative payoff. IR can be ensured by building upon the theoretical results from Myerson's system \cite{myerson}. It can also be enforced by integrating an additional constraint into the objective function and solve it using Augmented Lagrange Solver \cite{RegretNet, DLWithBudgets}

\subsection{Data Sparsity \& High Dimensionality}
The action space can blow up very quickly as the number of agents increase. The high dimensionality in action space can introduce severe computational burdens. In order to mitigate the computational burden, Tang et al. and Cai et al. decomposed the neural network into sub-networks for each seller and discretized the action spaces \cite{RLMechanismDesign, RLECommerce}.

In addition, many features are high-dimensional one-hot vectors, so data can be very sparse. The original regularization approaches take the entire vector into computations and the regularized vector has non-zero entries for most of the positions, which increases the computational time for sparse data drastically. Zhou et al. proposed a mini-batch aware regularization \cite{DeepInterestNetwork} approach, where only parameters of features appearing in the mini-batch participate in the computation of regularization.

\begin{equation} \label{eq:MiniBatch}
    w_j \leftarrow w_j - \eta [\frac{1}{|\mathcal{B}_m|} \sum_{(x, y) \in \mathcal{B}_m} \frac{\partial L(p(x), y)}{\partial w_j} + \lambda \frac{\max_{(x, y) \in \mathcal{B}_m} 1_{x_j \neq 0}}{n_j}w_j]
\end{equation}

The mini-batch aware regularization is shown in Equation \ref{eq:MiniBatch}, where $\eta$ is the learning rate, $\mathcal{B}_m$ is the $m^{th}$ batch, $w_j$ is the weight of $j^{th}$ feature, $n_j$ is the number of occurrences of $j^{th}$ feature. The numerator $\max_{(x, y) \in \mathcal{B}_m} 1_{x_j \neq 0}$ denotes if at least one instance in the $m^{th}$ mini-batch has feature $j$.

\subsection{Multiple Performance Metrics and Objectives}
The most intuitive objective for most auction designers is to maximize their profit. However, to adapt to the dynamic and complex nature of today's online auction systems, designers may be better-off if they consider multiple performance metrics. Zhang et al. listed out several commonly used performance metrics (See \ref{Metrics}). Zhang et al. and Liu et al. optimizes a linear combination of functions of those metrics, where both the linear weights and functions of those metrics and be specified by auction designers \cite{MultipleMetricsDeepGSP, NeuralAuction}. Over the time, designers are free to adjust the weights and functions if their objectives have changed.

\subsection{Fairness \& Fraud Prevention} \label{sec:Fairness}
Due to the biases in the training data, many online platforms have discriminatory ad allocations among different demography \cite{ProportionNet}. One of the major social problems associated with online advertising is the use in the job market, where unfairness can be very detrimental to the equality and the protection of underrepresented groups.

To mitigate the unfairness, PreferenceNet \cite{PreferenceNet} integrated three definitions of fairness into the model, all of which map the allocations $g(b)$ onto $\mathbb{R}$. In all these equations (Equation \ref{eq:TVF}, Equation \ref{eq:Entropy}, and Equation \ref{eq:Quota}), $i$ refers to $i^{th}$ item while $j$ and $j'$ refer to $j^{th}$ and $j'^{th}$ agents.
\begin{enumerate}
    \item Total Variation Fairness (Equation \ref{eq:TVF}): the distance between allocations cannot be larger than the discrepancies between these two users.
    \begin{equation} \label{eq:TVF}
        \sum_{i \in C_k} |g(b)_{i, j} - g(b)_{i, j'}| \leq d^k(j, j'), \forall k \in \{1, \dots, c\}, \forall j, j' \in M
    \end{equation} 
    \item Entropy (Equation \ref{eq:Entropy}): the allocation for an agent tends to be more uniformly distributed.
    \begin{equation} \label{eq:Entropy}
        \max -\sum_{i = 1}^n P(\frac{g(b)_{i.}}{\sum_j g(b)_{ij}})\log P(\frac{g(b)_{i.}}{\sum_j g(b)_{ij}})
    \end{equation}
    \item Quota (Equation \ref{eq:Quota}): the smallest allocation to any agent should be greater than some threshold.
    \begin{equation} \label{eq:Quota}
        \min_{j} (\frac{g(b)_{.j}}{\sum_{i} g(b)_{ij}}) > t
    \end{equation}
\end{enumerate}

ProportionNet \cite{ProportionNet} also adopted the notion of total variation fairness, so that the allocations to similar users cannot differ by too much. It converted the Equation \ref{eq:TVF} into an unfairness constraint (Equation \ref{eq:TVFConstraint}) that can be fed into the Augmented Lagrange Solver, which allows us to quantify the unfairness of the auction outcome for all users involved.
\begin{equation} \label{eq:TVFConstraint}
    unf_j = \sum_{j' \in M}\sum_{C_k \in C} \max(0, \sum_{i \in C_k}\max(0, z_{ij} - z_{ij'})) - d^k(j, j'))
\end{equation}

While unfairness can be introduced by the biases in auction mechanisms, it can also be induced by shill bidding behaviors in auctions. Sellers can adopt a variety of shill bidding strategies to inflate the final selling price of an item. The common four shill bidding strategies have been identified \cite{ShillBidding}:
\begin{enumerate}
    \item Evaluator: single bid engagement at an early time with a high amount.
    \item Sniping: single bid engagement in the last moment, not leaving opportunity for anybody else to outbid.
    \item Unmasking: multiple bid engagement in a short span of time with a probability of intend to exposing the maximum bid or the highest bidders.
    \item Skeptic: multiple bid engagement with lowest possible bids each time.
\end{enumerate}

\section{Conclusion}
In this article, we have gone through the rough evolving process of deep learning based online auction systems. Mechanisms designed using MLP infrastructure are usually built upon some theoretical results, and the MLP structure is used to represent the functions given in the theories. A more sophisticated structure came with the appearance of RegretNet, which parametrizes allocation rules and payment rules using separate networks. Many researchers have built extensions of RegretNet by integrating more constraints into the objective function or slightly adjusting the network structure but still keeping allocation and payment networks separate. The dynamic nature of online auction has encouraged researchers to adopt the deep reinforcement learning framework, which is more often a model-free approach that requires less assumptions on the data and is able to keep adapting itself as time progresses. As most traditional neural network has an implicit constraints on the positions of inputs, it integrates the ordering of auction participants into the model training, while in reality there is no inherent ordering among them. As a result, a deep learning based on DeepSet infrastructure has emerged, which can remove the effects of positions completely. We have also discussed the constraints and concerns faced by auction designers and we pointed out how researchers have attempted to address them.

Although researchers are progressing rapidly to the development of an online auction mechanism that can be reliable and profitable in the long term, there are still a lot of unresolved issues left for future researchers to investigate. As the size of users for online auction system is usually gigantic, the computational constraint and convergence problems for high-dimensional data are still non-negligible issues. Researchers are either mapping the data to lower dimension or reducing the action space by discretizing it, but it remains unclear how much information we are losing. In addition, most models assume the unit-demand and additive valuations, while this assumption might not be true in the real world. Last but not least, as most mechanism frameworks rely on the assumption that auction participants are independent from each other, their IC and IR constraints are also computed at individual levels. Therefore, their strategies might not be robust to the non-truthful behaviors conducted by participants in collusion.

\newpage

\begin{small}
\bibliographystyle{plain}
\bibliography{references}
\end{small}

\end{document}